\documentclass[10pt,journal,compsoc]{IEEEtran}
\usepackage{bbm}

\usepackage[T1]{fontenc}
\usepackage{amsmath,amssymb,amsfonts,mathrsfs,bm}
\usepackage{amsthm}
\usepackage{cite}
\usepackage{array}
\usepackage[shortlabels]{enumitem}
\usepackage{graphicx}
\usepackage{url}
\usepackage{color}
\usepackage{algorithm,algorithmic}
\usepackage{multirow}
\usepackage[table]{xcolor}
\usepackage[normalem]{ulem}
\usepackage{array}
\newcolumntype{L}[1]{>{\raggedright\let\newline\\\arraybackslash\hspace{0pt}}m{#1}}
\newcolumntype{C}[1]{>{\centering\let\newline\\\arraybackslash\hspace{0pt}}m{#1}}
\newcolumntype{R}[1]{>{\raggedleft\let\newline\\\arraybackslash\hspace{0pt}}m{#1}}
\usepackage{subfigure}
\usepackage{xparse}



\ifx\notloadhyperref\undefined
	\ifx\loadbibentry\undefined
		\usepackage[hidelinks]{hyperref} 
	\else
		\usepackage{bibentry}
		\makeatletter\let\saved@bibitem\@bibitem\makeatother
		\usepackage[hidelinks]{hyperref}
		\makeatletter\let\@bibitem\saved@bibitem\makeatother
	\fi
\else
\relax
\fi

\usepackage[capitalize]{cleveref}
\crefname{equation}{}{}
\Crefname{equation}{}{}
\crefname{claim}{claim}{claims}
\crefname{step}{step}{steps}
\crefname{line}{line}{lines}
\crefname{dmath}{}{}
\crefname{dseries}{}{}
\crefname{dgroup}{}{}
\crefname{Theorem}{Theorem}{Theorems}
\crefname{Corollary}{Corollary}{Corollaries}
\crefname{Proposition}{Proposition}{Propositions}
\crefname{Lemma}{Lemma}{Lemmas}
\crefname{Definition}{Definition}{Definitions}
\crefname{Example}{Example}{Examples}
\crefname{Assumption}{Assumption}{Assumptions}
\crefname{Remark}{Remark}{Remarks}
\crefname{Theorem_A}{Theorem}{Theorems}
\crefname{Corollary_A}{Corollary}{Corollaries}
\crefname{Proposition_A}{Proposition}{Propositions}
\crefname{Lemma_A}{Lemma}{Lemmas}
\crefname{Definition_A}{Definition}{Definitions}

\ifx\useTheoremCounter\undefined
\newtheorem{Theorem}{Theorem}
\newtheorem{Corollary}{Corollary}
\newtheorem{Proposition}{Proposition}
\newtheorem{Lemma}{Lemma}
\else
\newtheorem{Theorem}{Theorem}

\fi

\theoremstyle{remark}

\newcommand{\Nat}{\mathbb{N}}

\DeclareSymbolFont{bsfletters}{OT1}{cmss}{bx}{n}
\DeclareSymbolFont{ssfletters}{OT1}{cmss}{m}{n}
\DeclareMathSymbol{\bsfGamma}{0}{bsfletters}{'000}
\DeclareMathSymbol{\ssfGamma}{0}{ssfletters}{'000}
\DeclareMathSymbol{\bsfDelta}{0}{bsfletters}{'001}
\DeclareMathSymbol{\ssfDelta}{0}{ssfletters}{'001}
\DeclareMathSymbol{\bsfTheta}{0}{bsfletters}{'002}
\DeclareMathSymbol{\ssfTheta}{0}{ssfletters}{'002}
\DeclareMathSymbol{\bsfLambda}{0}{bsfletters}{'003}
\DeclareMathSymbol{\ssfLambda}{0}{ssfletters}{'003}
\DeclareMathSymbol{\bsfXi}{0}{bsfletters}{'004}
\DeclareMathSymbol{\ssfXi}{0}{ssfletters}{'004}
\DeclareMathSymbol{\bsfPi}{0}{bsfletters}{'005}
\DeclareMathSymbol{\ssfPi}{0}{ssfletters}{'005}
\DeclareMathSymbol{\bsfSigma}{0}{bsfletters}{'006}
\DeclareMathSymbol{\ssfSigma}{0}{ssfletters}{'006}
\DeclareMathSymbol{\bsfUpsilon}{0}{bsfletters}{'007}
\DeclareMathSymbol{\ssfUpsilon}{0}{ssfletters}{'007}
\DeclareMathSymbol{\bsfPhi}{0}{bsfletters}{'010}
\DeclareMathSymbol{\ssfPhi}{0}{ssfletters}{'010}
\DeclareMathSymbol{\bsfPsi}{0}{bsfletters}{'011}
\DeclareMathSymbol{\ssfPsi}{0}{ssfletters}{'011}
\DeclareMathSymbol{\bsfOmega}{0}{bsfletters}{'012}
\DeclareMathSymbol{\ssfOmega}{0}{ssfletters}{'012}

\DeclareMathOperator*{\argmax}{arg\,max}

\newcommand{\qednew}{\nobreak \ifvmode \relax \else
      \ifdim\lastskip<1.5em \hskip-\lastskip
      \hskip1.5em plus0em minus0.5em \fi \nobreak
      \vrule height0.75em width0.5em depth0.25em\fi}

\newcommand{\indicatore}[1]{{\bf 1}_{{#1}}}
\newcommand{\indicator}[1]{{\bf 1}_{\{{#1}\}}}

\newcommand{\ofrac}[1]{{\frac{1}{#1}}}

\newcommand{\ceil}[1]{\left\lceil{#1}\right\rceil}

\newcommand{\cond}[2]{\left. {#1}\, \middle| \, {#2} \right.}

\DeclareDocumentCommand \P { g d() g } {%
	\IfNoValueTF {#3} 
	{%
		\IfNoValueTF {#1} 
		{%
			\IfNoValueTF {#2}
			{%
				\mathbb{P}%
			}%
			{%
				\mathbb{P}\left({#2}\right)%
			}%
		}%
		{%
			\IfNoValueTF {#2}
			{%
				\mathbb{P}_{#1}%
			}%
			{%
				\mathbb{P}_{#1}\left({#2}\right)%
			}%
		}%
	}%
	{%
		\IfNoValueTF {#1} 
		{%
			\mathbb{P}\left(\cond{#2}{#3}\right)%
		}%
		{%
			\mathbb{P}_{#1}\left(\cond{#2}{#3}\right)%
		}%
	}%
}

\DeclareDocumentCommand \E { g o g } {%
	\IfNoValueTF {#3} 
	{%
		\IfNoValueTF {#1} 
		{%
			\IfNoValueTF {#2}
			{%
				\mathbb{E}%
			}%
			{%
				\mathbb{E}\left[{#2}\right]%
			}%
		}%
		{%
			\IfNoValueTF {#2}
			{%
				\mathbb{E}_{#1}%
			}%
			{%
				\mathbb{E}_{#1}\left[{#2}\right]%
			}%
		}%
	}%
	{%
		\IfNoValueTF {#1} 
		{%
			\mathbb{E}\left[\cond{#2}{#3}\right]%
		}%
		{%
			\mathbb{E}_{#1}\left[\cond{#2}{#3}\right]%
		}%
	}%
}

\newcommand{\Bern}[1]{\mathrm{Bern}\left(#1\right)}

\definecolor{gray90}{gray}{0.9}

\newcommand{\msout}[1]{\text{\color{green} \sout{\ensuremath{#1}}}}
\newcommand{\del}[1]{{\color{green}\ifmmode \msout{#1}\else\sout{#1}\fi}}

\newcommand{\hide}[1]{}

\renewcommand{\figurename}{Fig.}

\graphicspath{{./Figures/}} 
\begin{document}

\title{Task Recommendation in Crowdsourcing Based on Learning Preferences and Reliabilities}
\author{Qiyu Kang, \IEEEmembership{Student Member,~IEEE,} and Wee Peng Tay, \IEEEmembership{Senior Member,~IEEE}

\thanks{This research is supported in part by the Singapore Ministry of Education Academic Research Fund Tier 2 grant MOE2014-T2-1-028.}
\thanks{The authors are with the School of Electrical and Electronic Engineering, Nanyang Technological University, Singapore. Email: KANG0080@e.ntu.edu.sg, wptay@ntu.edu.sg.}  }

\IEEEtitleabstractindextext{
\begin{abstract}
Workers participating in a crowdsourcing platform can have a wide range of abilities and interests. An important problem in crowdsourcing is the task recommendation problem, in which tasks that best match a particular worker's preferences and reliabilities are recommended to that worker. A task recommendation scheme that assigns tasks more likely to be accepted by a worker who is more likely to complete it reliably results in better performance for the task requester. Without prior information about a worker, his preferences and reliabilities need to be learned over time. In this paper, we propose a multi-armed bandit (MAB) framework to learn a worker's preferences and his reliabilities for different categories of tasks. However, unlike the classical MAB problem, the reward from the worker's completion of a task is unobservable. We therefore include the use of gold tasks (i.e., tasks whose solutions are known \emph{a priori} and which do not produce any rewards) in our task recommendation procedure. Our model could be viewed as a new variant of MAB, in which the random rewards can only be observed at those time steps where gold tasks are used, and the accuracy of estimating the expected reward of recommending a task to a worker depends on the number of gold tasks used. We show that the optimal regret is $O(\sqrt{n})$, where $n$ is the number of tasks recommended to the worker. We develop three task recommendation strategies to determine the number of gold tasks for different task categories, and show that they are order optimal. Simulations verify the efficiency of our approaches.  
\end{abstract}

}
\maketitle

\IEEEdisplaynontitleabstractindextext
\IEEEpeerreviewmaketitle

\IEEEraisesectionheading{\section{Introduction}\label{sec:intro}}

\IEEEPARstart{I}{n} a typical crowdsourcing platform, a worker may be recommended a wide variety of tasks to choose from \cite{CrowdFlower,Amazoncompany,crowdspring}. For example, on Amazon Mechanical Turk (MTurk) and CrowdFlower, tasks can include labeling the content of an image, determining whether or not sentences extracted from movie reviews are positive, or determining whether or not a website has adult contents. In these examples, the different tasks require different sets of skills from a worker. In the image labeling task, a worker who is good at visual recognition will perform reliably, whereas identifying whether a movie review is positive or not requires a worker with language skills and knowledge of the nuances of that particular language the review is written in.  Workers may also have different interests, and may choose not to accept tasks that they are qualified for. Therefore, the crowdsourcing platform needs to find a way to best match the available tasks to the most suitable workers to improve the likelihood of obtaining high-quality solutions to the tasks.

In this paper, we assume that tasks are organized into several categories. Tasks belonging to the same category need similar domain knowledge to be completed. Within the task assignment context, a number of studies have focused on offline strategies, which assign the tasks to workers without any adaptation or feedback during the workers' tasks completion process. For example, in \cite{Karger2014, karger2013efficient}, the authors proposed a task assignment scheme using a bipartite graph to model the affinity of workers for different binary tasks and an iterative algorithm based on belief propagation to infer the final decision from the workers' responses. Reference \cite{basik2018fair} studied a task allocation model that provides distributional fairness among workers, while maximizing the task allocation ratio. A privacy-preserving task recommendation scheme was proposed in \cite{shu2018privacy}, which only considers workers' preferences without taking workers' reliabilities into account.
The paper \cite{khetan2016reliable} considered heterogeneous tasks with different difficulties using the generalized Dawid-Skene model \cite{zhou2015regularized}, while \cite{cappe2009line} utilizes the expectation maximization approach to estimate the tasks' solutions and workers' reliabilities. The aforementioned approaches are all offline, i.e., the task assignments are made without taking into account additional information about the workers' performances and behaviors that can be gleaned while the workers complete tasks sequentially over time.

Online learning for adaptive task assignments has also been considered in recent years. For instance, in \cite{kang2017sequential,kang2017}, as workers may be unreliable, the authors proposed to perform sequential questioning in which the questions posed to the workers are designed based on previous questions and answers. Reference \cite{massoulie2015greedy} studied sequential user selection using a Bayes update aided online solution in the problem of news disseminating. The paper \cite{ho2013adaptive} investigated the problem of label inference for heterogeneous classification tasks by applying online primal-dual techniques. In \cite{ho2012online}, the authors studied sequential task assignment with budgets that specify how many times the requester would like each task to be completed. Expert-crowdsourcing in which workers are experts who have unknown reliabilities was studied in \cite{tran2014efficient}, which also assumed that the requester has a budget constraint and each worker has a maximum task limit. The authors proposed an algorithm, which uses an initial exploration phase to uniformly sample the performance of a wide range of workers, and in the latter exploitation phase solves a bounded knapsack problem \cite{kellerer2003knapsack}. In the mobile micro-task allocation problem in spatial crowdsourcing scenarios, where the orders of arriving tasks and workers are dynamic, \cite{tong2016online} proposed an online two-phase-based bipartite graph matching algorithm with good average performance. In \cite{fan2015icrowd}, the authors proposed an adaptive crowdsourcing framework, which continuously estimates the reliability of a worker by evaluating her performance on the completed tasks, and predicts which tasks the worker is well acquainted with. In the above work, the authors assumed that the crowdsourcing platform is able to quickly and accurately evaluate the quality of all the tasks that have been completed. However, as is often the case in practice, the crowdsourcing platform cannot directly and immediately evaluate the quality of a completed task \cite{nee2018context}. For example, in tasks where workers are asked to determine whether an image contains a particular object, the crowdsourcing platform has no way to determine if the task completed by a worker is successful or not. In \cite{liu2017online}, the authors proposed an online algorithm, in which  workers' reliabilities are estimated by repetitively assigning each worker the same tasks occasionally.

The above-mentioned works do not take workers' interests or preferences for certain tasks into consideration when performing a task assignment. In most existing crowdsourcing platforms like MTurk, workers may choose to accept or reject a recommended task. As the number of tasks uploaded by requesters daily can be large \cite{safran2017real}, the crowdsourcing platform should aim to recommend tasks to each worker that he will likely accept in order to optimize the platform productivity. Including the workers' preferences in task assignment can also help workers to find their preferred tasks faster as they will have fewer recommendations to consider. At the same time, by estimating the reliability of a worker in each task category helps the requester to collect high-quality results quicker \cite{yuen2012task}. The paper \cite{chen2010sjc} proposed to perform task assignment using workers' preferences based on many-to-one matching. However, the paper assumes that workers' preferences and reliabilities are known \emph{a priori}, which is unrealistic in a practical system. In \cite{safran2017real}, the authors proposed two recommendation strategies that compute the top-$k$ most suitable tasks for a given worker and the top-$k$ best workers for a given task. However, they assume that the workers can be evaluated immediately. The references \cite{yuen2015taskrec, yuen2012task} employed matrix factorization techniques to capture workers' preferences and reliability where, however, the evaluated scores in the matrix are also assumed to be gotten immediately. 

In this paper, we propose an online learning approach to estimate workers' preferences and reliabilities. We assume that tasks can be categorized, and that each worker can choose to accept or reject a recommended task at each time step. If he chooses to accept the task, he completes it within the current time step. For each time step, our goal is to recommend a task from a given number of task categories so that it best matches the worker's preferences and reliabilities according to a reward function. As the worker's preferences and reliabilities for different task categories are unknown \emph{a priori}, we adopt a multi-armed bandit (MAB) formulation \cite{lai1985asymptotically, auer2002finite} in which these are learnt through exploration, while the cumulative sum reward is optimized through the exploitation of the empirical best match. However, unlike the classical MAB problem in which the reward at each time step can be observed, the reward is \emph{unobservable} in our formulation as the crowdsourcing platform does not know the true solution of each task. To overcome this, we include \emph{gold} tasks \cite{kittur2008crowdsourcing} whose solutions are known \emph{a priori}. In \cite{le2010ensuring, oleson2011programmatic}, gold tasks are used for training and estimating workers' reliabilities. However, gold tasks do not generate any rewards. Including more gold tasks leads to better estimates of the worker's preferences and reliabilities for different task categories, but also results in less reward. In a crowdsourcing platform, a plausible compensation scheme is to reward a worker for completed tasks according to his reliability \cite{geiger2011managing,kaufmann2011more, jarrett2018crowdsourcing}. Furthermore, to accurately fuse the results from multiple workers, the knowledge of the workers' reliabilities are required \cite{ho2013adaptive,li2014error}.  Therefore, the platform aims to estimate the worker's reliability as accurately as possible. To reflect this, we include both the reward and estimation accuracy in computing the regret. We show that the optimal regret under our formulation has order $O(\sqrt{n})$, where $n$ is the number of time steps. In addition, we propose three strategies that achieve the order optimal regret. 

Our MAB formulation is related to, but different from, the risk-averse MAB problem considered in \cite{vakili2016risk,vakili2015mean,sani2012risk}, which uses mean-variance \cite{steinbach2001markowitz, markowitz1952portfolio} as its risk objective. The differences between our MAB formulation and the risk-averse MAB are: (i) The worker in our problem may choose not to accept a task at each time step, leading to no reward at that time step, whereas in the risk-averse MAB, such an option is not available. (ii) The variance of the reward is used in the formulation of the risk-averse MAB, while we use the variance of the reward \emph{estimator} in formulating our regret. This is because as explained above, in a crowdsourcing platform, the reward at each time step is unobservable, and our regret is dependent on the estimation accuracy from using gold tasks to estimate the expected reward instead of the reward variance.      

The rest of this paper is organized as follows. In \cref{sec:model}, we present our system model and assumptions. In \cref{sec:algor}, we derive the optimal order of the regret and introduce three strategies that are order optimal in \cref{sec:strategies}. In \cref{sec:simulation}, we present simulations to compare the performance of our approaches. \Cref{sec:conclusion} concludes the paper.

\emph{Notations:} We use $\Bern{p}$ to denote the distribution of a Bernoulli random variable $X$ with $\P(X=1)=p$. The notation $\sim$ denotes equality in distribution. We use $\Nat$ to denote the set of positive integers and $E^c$ to denote the complement of the event $E$. The indicator function $\indicatore{A}(\omega)=1$ if and only if $\omega\in A$. We let $x^+ = \max\{x,0\}$. For non-negative functions $f(n)$ and $g(n)$, we write $f(n)=O(g(n))$ if $\limsup_{n\to\infty} f(n)/g(n)<\infty$, $f(n) = o(g(n))$ if $\lim_{n\to\infty} f(n)/g(n)=0$, and $f(n) = \Theta(g(n))$ if $0 < \liminf_{n\to\infty} f(n)/g(n) \leq \limsup_{n\to\infty} f(n)/g(n)<\infty$.

\section{System Model}
\label{sec:model}
We consider a crowdsourcing platform where each task belongs to one of $K$ categories. The platform recommends a task from some category to a worker at each time $n\in \Nat$. The worker has different reliabilities and preferences for different categories. The worker's reliability for category $k$ is the probability $p_k$ of completing a task from category $k$ correctly, and his preference refers to the probability $q_k$ of accepting a task from category $k$. We let $A_{k,t} = 1$ if the worker accepts the $t$-th task recommended to him from category $k$, and $A_{k,t} = 0$ otherwise. We let $X_{k,t}=1$ if the worker completes the task correctly, and $X_{k,t} = 0$ otherwise. For all $k$ and $t$, we assume that $A_{k,t}\sim \text{Bern}(q_k)$ are independent and identically distributed. Similarly, $X_{k,t}\sim \text{Bern}(p_k)$ are independent and identically distributed. We assume that $A_{k,t}$ and $X_{k,t}$ are independent. 

We model the task recommendation problem as a MAB problem. However, since $X_{k,t}$ is typically unobservable since the system does not know the task solution \emph{a priori}, we include the use of gold tasks in our recommendation. A gold task is a task whose solution is known \emph{a priori} to the system. A reward is obtained only if the worker accepts and completes a non-gold task correctly.

Let $Y_{k,t}=A_{k,t}X_{k,t}$, whose distribution is $\text{Bern}(q_k p_k)$. We denote the task category that maximizes $q_k p_k$ as $* = \argmax_{1\le k \le K} q_k p_k$. For each time step $t$, let $G_{k,t}$ denote the event that the $t$-th task from category $k$ recommended to the worker is a gold task. Let
\begin{align}
g_k(t) = \sum_{i=1}^t A_{k,i}\indicatore{G_{k,i}} 
\end{align}
be the number of completed gold tasks. If the worker completes a non-gold task, he is rewarded in proportion to $p_k$, the probability that he has completed the task correctly. However, since we do not know $p_k$ \emph{a priori}, the crowdsourcing platform estimates it using the empirical mean from the $g_k(t)$ completed gold tasks: 
\begin{align}\label{barX}
\bar{X}_{k}(t) = \ofrac{g_k(t)}  \sum_{i=1}^{t} A_{k,j_i} X_{k,j_i} \indicatore{G_{k,i}}.
\end{align}
Due to the uncertainty in the estimate \cref{barX}, we assume that the reward for the $t$-th non-gold task from category $k$ if the worker completes it is given by
\begin{align}\label{beta}
\left(p_k - \beta \frac{\sigma_k^2}{g_k(t)}\right)^{+},
\end{align}%
where $\beta >0$ is a predefined weight that quantifies the importance of the estimator $\bar{X}_{k}(t)$'s accuracy to the crowdsourcing platform, and $\sigma_k^2=p_k(1-p_k)$ is the variance of a task reward from category $k$. The quantity $\sigma_k^2/g_k(t)$ is the variance of $\bar{X}_k(t)$ conditioned on $g_k(t)$. 

For each $k=1,\ldots,K$ and time $n$, let $N_k(n)$ denote the number of tasks recommended so far till time $n$ from category $k$. We define the cumulative reward function at time $n$ as:
\begin{align}
	r(n)
	&= \E \sum_{k=1}^{K}\sum_{t=1}^{N_k(n)} A_{k,t}\left( X_{k,t} -\beta \frac{\sigma_k^2}{g_k(t)} \right)^{+}\indicatore{G_{k,t}^c}\nonumber\\
	&= \E \sum_{k=1}^{K}\sum_{t=1}^{N_k(n)} q_k\left( p_k -\beta \frac{\sigma_k^2}{g_k(t)} \right)^{+}\indicatore{G_{k,t}^c},
	\label{reward}
\end{align}
since $A_{k,t}$, $X_{k,t}$ and $(G_{k,t},N_k(n))$ are independent. Note that the coefficient $\beta$ can also be interpreted as the Lagrange multiplier for a constrained optimization problem in which the reward
\begin{align*}
\E \sum_{k=1}^{K}\sum_{t=1}^{N_k(n)} q_{k} p_{k}\indicatore{G_{k,t}^c}
\end{align*}
is maximized subject to a constraint on the uncertainty in the estimator \cref{barX}.

To avoid dividing by zero, i.e., to ensure $g_k(t) > 0$, we assume that the worker completes one gold task from each category before the recommendation starts, which can be done through a calibration process in a crowdsourcing platform.

To maximize the above reward function \eqref{reward} is equivalent to minimizing the regret function at time $n$, defined as:
\begin{align}
	R(n) = nq_*p_* - r(n). \label{regret}
\end{align}

For simplicity, we assume that every task category is non-empty at each time step. This is a reasonable assumption since the task pool in a crowdsourcing platform is updated constantly, and a single task may be assigned to more than one worker. In the following, we use the terms ``task category'' and ``arm'' interchangeably in our MAB formulation.

\section{Optimal Regret Order}
\label{sec:algor}
In this section, we first show that the optimal order of the regret function \eqref{regret} is $O(\sqrt{n})$, where $n$ is the number of time steps. We then propose three recommendation strategies, all of which achieve the order optimal regret.
\begin{Theorem}\label{theorem:1}
	The optimal order of the regret function \eqref{regret} is $O(\sqrt{n})$, where $n$ is the number of time steps.
\end{Theorem}
\begin{IEEEproof}
Let 
\begin{align}
f(n) = \sum_{k=1}^K \sum_{t=1}^{N_k(n)} \indicatore{G_{k,t}} \label{fn}
\end{align}
be the total number of gold tasks recommended till time $n$. Then, from \cref{reward}, we have
\begin{align} 
r(n) & = \E \sum_{k=1}^{K}\sum_{t=1}^{N_k(n)} q_k\left( p_k -\beta \frac{\sigma_k^2}{g_k(t)} \right)^{+}\indicatore{G_{k,t}^c}\nonumber\\
& \le \E \sum_{k=1}^{K}\sum_{t=1}^{N_k(n)} \left(q_* p_* -\beta \frac{q_k\sigma_k^2}{f(n)} \right)^{+}\indicatore{G_{k,t}^c}\nonumber\\
	& \le \E[(n-f(n))\left(q_*p_* - \beta \frac{\min_kq_k\sigma_k^2}{f(n)}\right)]^{+}.\label{rbdd}
\end{align}
From \cref{regret,rbdd}, with $a=\beta \min_kq_k\sigma_{k}^{2}$, when $n$ is sufficiently large, we then have
\begin{align}
	R(n) &\ge nq_*p_* - \E[(n-f(n))\left(q_*p_* - \beta \frac{\min_kq_k\sigma_k^2}{f(n)}\right)]^{+}\nonumber\\
	&\ge \E \left[\min \left\{ a\frac{n}{f(n)}+ q_*p_*f(n) - a , nq_*p_*\right\} \right] \label{Rbdd1}\\
	&\ge \min \left\{ 2 \sqrt{a q_*p_* n} - a, nq_*p_*\right\} \label{Rbdd2} \\
		& = 2 \sqrt{a q_*p_* n} - a, 
\end{align}
where the inequality in \cref{Rbdd2} follows because $a\frac{n}{f(n)}+ q_*p_*f(n) - a \geq 2 \sqrt{a q_*p_* n} - a$ with probability $1$. The theorem is now proved.
\end{IEEEproof}

\section{Order Optimal Strategies}\label{sec:strategies}

In this section, we propose three strategies that achieve the optimal order regret of $O(\sqrt{n})$ in \cref{theorem:1}, and discuss the advantages of each.

\subsection{Greedy Recommendation Strategy}

In our greedy recommendation strategy (GR), we divide time into epochs. In each epoch $r$, a single task category or arm $c_r$ is chosen and all tasks in that epoch are drawn from that category. In the first $K$ epochs, each of which consists of a single time step, the worker completes a gold task from each of the $K$ categories. Subsequently, the $r$-th epoch, where $r\geq K+1$, consists of $\tau(r) - \tau(r-1)+1$ time steps, where $\tau(r) =  \lceil\alpha r^{2}\rceil$, and $\alpha>0$. In each of these epochs, we set the first task to be a gold task, while for all other time steps in the epoch, the tasks recommended are non-gold tasks chosen from a particular task category. For each $r\geq K+1$, the task category $c_r$ is chosen based on a $\epsilon_r$-greedy policy \cite{auer2002finite} as follows.  

Let $\Delta_k = q_*p_*-q_kp_k$, and choose $d$ so that $0< d\le \min_{k\ne *} \Delta_k$. For each $r = 1,2,\ldots$, let $\epsilon_r = \min \{1, cK/(d^2r)\}$, where $c>0$ is a fixed constant. At the beginning of each epoch $r$, where $r \geq K+1$, we choose 
\begin{align}
z_r = \argmax_k \bar{Y}_k(T_k(r-1)),\label{zr}
\end{align}
where $T_k(r)$ is the number of epochs within the first $r$ epochs that category $k$ is chosen, 
\begin{align}
\bar{Y}_k(g) = \ofrac{g} \sum_{i=1}^{g} A_{k,j_i} X_{k,j_i},
\end{align} 
and $j_1,\ldots,j_g$ are the indices of the first $g$ gold tasks recommended from category $k$. Then, with probability $1-\epsilon_r$, we let $c_r = z_r$ to be the task category chosen for epoch $r$, and with probability $\epsilon_r$, we let $c_r$ to be a randomly chosen category. The GR strategy is summarized in \cref{algor1}.

\begin{algorithm}[!htb]
\caption{GR}
\label{algor1} 
\begin{algorithmic}[1]
\REQUIRE $0< d \le \min_{k\ne *} \Delta_k$, $c > 0$, $\alpha>0$.
\ENSURE 
\STATE Recommend one gold task from each category. 
\STATE Set $r = K+1$, $T_k(r) = 1$ and $\bar{Y}_{k}(1) = Y_{k,1}$ for all $k=1,\ldots,K$.%
\LOOP
\STATE Find $z_r$ in \cref{zr}.%
\STATE Set $\epsilon_r = \min \{1, \frac{cK}{d^2r}\}$. With probability $1-\epsilon_r$, choose $c_r = z_r$, otherwise set $c_r$ to be a randomly chosen arm.%
\STATE Recommend one gold task from arm $c_r$ at the first time step of epoch $r$. Update $T_{c_r}(r)=T_{c_r}(r-1)+1$, $n_{c_{r}}(r)$ and $\bar{Y}_{c_r}(T_{c_r}(r))$.%
\STATE Recommend a non-gold task drawn from arm $c_r$ at each of the following $\tau(r) - \tau(r-1)$ time steps.%
\STATE Set $r = r + 1$.
\ENDLOOP
\end{algorithmic} 
\end{algorithm}

\begin{Theorem}\label{theorem:GR}
Suppose that $c >\max\{5d^2, 2\}$. Then, GR has regret of order $O(\sqrt{n})$, where $n$ is the number of time steps. 
\end{Theorem}
\begin{IEEEproof}
For each arm $k$ in epoch $r$, we denote by $P_{k,r} = \P(c_r = k)$, the probability that arm $k$ is chosen. From Theorem 3 in \cite{auer2002finite}, if $c > \max\{5d^2, 2\}$ and $k\ne *$, we have
\begin{align}\label{P_k_r}
	P_{k,r} \le \frac{c}{d^2r} + o(r^{-1}),
\end{align}
so that	
\begin{align}\label{P_star_r}
P_{*,r} \ge 1- \frac{c(K-1)}{d^2r}- o(r^{-1}).
\end{align}
Let $M$ be the number of epochs completed till the time step $n$, i.e., $M$ is the largest integer satisfying $ K+\sum_{r=K+1}^{M} \left( \tau(r) - \tau(r-1)+1\right)  = \lceil\alpha M^{2}\rceil- \lceil\alpha K^{2}\rceil + M \le n$. We have $M = O(\sqrt{n})$. Let $n_k(r)$ be the number of category $k$ gold tasks completed in the first $r$ epochs. Since a single gold question is recommended in each epoch, we have $g_{c_r}(t)=n_{c_r}(r)$ for all time steps $t$ in epoch $r$. From \cref{reward}, we obtain
\begin{align}
&r(n) \nonumber\\ 
&\geq  \E \sum_{r=K+1}^{M}q_{c_r}\left(\tau(r) - \tau(r-1)\right)\left(p_{c_r} -  \frac{\beta\sigma_{c_r}^{ 2}}{n_{c_r}(r)}\right)^{+} \nonumber \\	
&\ge \E \sum_{r=K+1}^{M}q_*(2\alpha r-\alpha-1)^{+}\left(p_* - \frac{\beta\sigma_{*}^{ 2}}{n_*(r)}\right)\indicator{c_r = *} \nonumber \\
&=	\sum_{r=K+1}^{M}q_*(2\alpha r-\alpha-1)^{+}\left(p_*P_{*,r} -  \E[\frac{\beta \sigma_{*}^{ 2}}{n_{*}(r)}]\right).\label{rn}
\end{align}
Recall also that $T_k(r) = \sum_{i=1}^{r} \indicator{c_i=k}$ is the number of epochs within the first $r$ epochs in which arm $k$ was chosen. Since GR assumes that the worker completes a gold task from arm $*$ before the recommendation starts, we have
\begin{align*}
n_{*}(r)-1 \mid T_*(r) \sim \Bern{T_*(r)-1,q_*},
\end{align*}
and
\begin{align} \label{bbb}
&\E[\ofrac{n_{*}(r)}] \nonumber\\
&=\E[\E[\ofrac{n_{*}(r)}]{T_*(r)}]\nonumber  \\
&= \E\sum_{j=0}^{T_*(r)-1}\ofrac{1+j}\binom{T_*(r)-1}{j}q_{*}^{j}(1-q_*)^{T_*(r)-1-j}\nonumber \\
&= \E\sum_{j=0}^{T_*(r)-1} \ofrac{T_*(r)} \binom{T_*(r)}{j+1}q_{*}^{j}(1-q_*)^{T_*(r)-1-j}\nonumber\\
&= \E\ofrac{T_*(r)q_*}\sum_{j=0}^{T_*(r)-1}\binom{T_*(r)}{j+1}q_{*}^{j+1}(1-q_*)^{T_*(r)-1-j}\nonumber\\
&= \E\ofrac{T_*(r)q_*}\left( 1-(1-q_*)^{T_*(r)}\right)\nonumber\\
&\le  \E[\ofrac{T_*(r)q_*}].
\end{align}
From \eqref{P_star_r}, \eqref{rn} and \eqref{bbb}, we obtain 
\begin{align}\label{rn2}
&r(n) \nonumber \\
&\ge \sum_{r=K+1}^{M}q_* p_*(2\alpha r-\alpha-1)^{+} \left( 1- O(r^{-1})\right) \nonumber \\
& \hspace{0.5cm} - \sum_{r=K+1}^{M}(2\alpha r-\alpha-1)^{+}\E  \left[\frac{\beta \sigma_{*}^{ 2}}{T_*(r)}\right]\nonumber \\
&\ge q_* p_* \alpha M^2 - O(M) - \sum_{r=K+1}^{M}(2\alpha r-\alpha-1)^{+}\E  \left[\frac{\beta \sigma_{*}^{ 2}}{T_*(r)}\right] \nonumber \\
&\ge nq_*p_* - O(\sqrt{n}) -\sum_{r=K+1}^{M}(2\alpha r-\alpha-1)^{+}\E  \left[\frac{\beta \sigma_{*}^{ 2}}{T_*(r)}\right].
\end{align} 
We next prove the following lemma.
\begin{Lemma}\label{lem:T*}
$\E[\ofrac{T_*(r)}] \le O(r^{-1})$.
\end{Lemma}
We have
\begin{align}\label{E_T_star1}
&\E[\ofrac{T_*(r)}] \nonumber\\
&=  \E[\ofrac{T_*(r)}]{T_*(r)\le \frac{r}{K}}\P(T_*(r)\le \frac{r}{K}) \nonumber\\
&\hspace{0.5cm} +\E[\ofrac{T_*(r)}]{T_*(r)> \frac{r}{K}}\P(T_*(r)> \frac{r}{K})\nonumber\\
&\le \P(T_*(r)\le \frac{r}{K}) + O(r^{-1}).
\end{align}
Therefore, to show the lemma, it suffices to show that the tail probability $\P(T_*(r)\le \frac{r}{K})$ is of order $O(r^{-1})$.

Consider $k\ne *$. Let $T_k^{B}(r)$ be the number of times $z_j=k$ for $j \leq r$, and $T_k^R(r)$ be the number of times arm $k$ was randomly chosen for $j \leq r$. From the union bound, we have
\begin{align} \label{T_k}
\P(T_k(r)\ge \frac{r}{K})\le \P(T_k^R(r)\ge \frac{r}{2K}) +\P(T_k^{B}(r) \geq \frac{r}{2K}).
\end{align}
From Hoeffding's inequality, we obtain
\begin{align}
\P(T_k^R(r)\ge \frac{r}{2K}) 
&\le e^{-2(\frac{1}{2K} - \E T_k^R(r)/r)^2r}\nonumber\\
&\le e^{-2(\frac{1}{2K} - O(\frac{c\ln r}{d^2r}))^2r}  \nonumber \\
&=  e^{-O(r)}, \label{P_T_k_R}
\end{align}
where the second inequality follows because the probability of randomly choosing arm $k$ at each epoch $j$ is not more than $cK/(d^2j) \cdot 1/K = c /(d^2j)$ and 
\begin{align*}
\E T_k^R(r) \le \sum_{j=1}^r \frac{c}{d^2 j} \le \frac{c (\ln r + 1)}{d^2}.
\end{align*}
In order to bound the second term in \eqref{T_k}, we use a similar argument as that in \cite{audibert2009exploration}. Let $u(r) = \frac{c}{d^2}\ln \frac{rd^2e^{1/2}}{2cK^2}$ and $s(r) = \ceil{r/(2K)}$. Define the following events:
\begin{align*}
E_1 &= \left\{\bar{Y}_{k}(j) \leq q_kp_k + \frac{\Delta_k}{2}, \text{ for all } j \in \left[s(r),r\right] \right\},\\
E_2 &= \left\{\bar{Y}_{*}(j) \ge q_*p_* - \frac{\Delta_k}{2}, \text{ for all } j \in (u(r), r] \right\},\\
E_3 &= \left\{T_*(s(r)) > u(r) \right\}.
\end{align*}
Under the event $E_1 \cap E_2 \cap E_3$, we have for all $i\in [s(r),r]$,
\begin{align*}
r \geq T_*(i) \geq T_*(s(r)) > u(r),
\end{align*}
which implies that
\begin{align}
\bar{Y}_*(T_*(i)) 
&\geq q_*p_* - \frac{\Delta_k}{2} \nonumber\\
&> q_kp_k + \frac{\Delta_k}{2} \nonumber\\
&\geq \bar{Y}_k(j), \label{Xkbd}
\end{align}
for all $j \in [s(r),r]$. Since $T_k^B(s(r))\leq s(r)$, we have $T_k^B(r) \leq s(r)$ because otherwise there exists $r'$ such $T_k^B(r')\in [s(r),r]$ and $\bar{Y}_k(T_k(r')) > \bar{Y}_*(T_*(r'))$, which contradicts \cref{Xkbd}. Therefore, $E_1 \cap E_2 \cap E_3 \subset \{T_k^B(r) \leq s(r)\}$, and 
\begin{align}
&\P(T_k^{B}(r)> s(r)) \nonumber\\
&\leq \P(E_1^c \cup E_2^c \cup E_3^c) \nonumber\\
&\le \P(E_1^c) + \P(E_2^c) + \P(E_3^c), \label{t_B0}
\end{align}
where the last inequality is the union bound. We next bound each of the terms on the right hand side of \cref{t_B0} separately. For the first term, we have
\begin{align}
\P(E_1^c)
&\leq \sum_{j = s(r)}^{r} \P(\bar{Y}_{k}(j) > q_kp_k + \frac{\Delta_k}{2}) \nonumber\\
&\leq \sum_{j = s(r)}^{r} e^{-\frac{\Delta_k^2}{2} j} \nonumber\\
&\leq \frac{e^{-\frac{\Delta_k^2 r}{4K}}}{1-e^{-\frac{\Delta_k^2}{2}}} \nonumber\\
&= o(r^{-1}), \label{t_B1}
\end{align}
where the second inequality follows from Hoeffding's inequality. Similarly, for the second term on the right hand side of \cref{t_B0}, we obtain
\begin{align}
\P(E_2^c)
&\leq \sum_{j = u(r)}^{r} \P(\bar{Y}_*(j) < q_*p_* -\frac{\Delta_k}{2}) \nonumber\\
&\leq \sum_{j = u(r)}^{r} e^{-\frac{\Delta_k^2}{2} j} \nonumber\\
&= O(e^{-\frac{\Delta_k^2c}{2d^2} \ln r}) \nonumber\\
&= o(r^{-1}),\label{t_B2}
\end{align}
since $c > 2$ and $0< d\le \min_{k\ne *} \Delta_k$. 

From the proof of Theorem 3 in \cite{auer2002finite}, we have $u(r) \le \frac{1}{2K}\sum_{i=1}^{s(r)}\epsilon_i$ and
\begin{align}
\P(E_3^c)
&\leq \P(T^R_*(s(r)) \le u(r)) \nonumber\\
&\leq \P(T^R_*(s(r)) \le \frac{1}{2K}\sum_{i=1}^{s(r)}\epsilon_i)\nonumber\\
&\leq e^{-\frac{1}{5}\frac{1}{2K}\sum_{i=1}^{s(r)}\epsilon_i} \nonumber\\
&\leq e^{-u(r)/5} \nonumber\\
&= \left(\frac{2cK^2}{r d^2 e^{1/2}}\right)^{\frac{c}{5d^2}} \nonumber\\
&= o(r^{-1}), \label{t_Br}
\end{align}
where the third inequality follows from Bernstein's inequality (see (13) of \cite{auer2002finite} for a detailed proof).

Applying \cref{t_B1,t_B2,t_Br} on the right hand side of \cref{t_B0}, we obtain $\P(T_k^B(r) > s(r)) = o(r^{-1})$. From \cref{P_T_k_R,T_k}, we then have
\begin{align*}
\P( T_k(r)\ge \frac{r}{K}) = o(r^{-1}),
\end{align*}
which yields

\begin{align}\label{P_T_star}
\P(T_*(r)\leq \frac{r}{K}) 
&= \P(\sum_{k\ne *} T_k(r)\ge \frac{r(K-1)}{K})\nonumber\\
&\leq \sum_{k\ne *} \P( T_k(r)\ge \frac{r}{K}) \nonumber\\
&= o(r^{-1}), 
\end{align}
and \cref{lem:T*} is proved.

From \cref{rn2,lem:T*}, we obtain 
\begin{align*}
	r(n) 
	&\ge nq_*p_* - O(\sqrt{n}) -\sum_{r=K+1}^{M}(2\alpha r-\alpha-1)^{+}O(r^{-1}) \nonumber \\ 
	&\ge np_*q_* - O(\sqrt{n}) - O(M) \nonumber \\ 
	&\ge np_*q_* - O(\sqrt{n}),
\end{align*}
since $M = O(\sqrt{n})$. The proof of \cref{theorem:GR} is now complete.
\end{IEEEproof}

\subsection{Uniform-Pulling Recommendation Strategy}

\begin{algorithm}[!ht]
\caption{UR}
\label{alg:UR} 
\begin{algorithmic}[1] 
\REQUIRE $\alpha>0$.
\ENSURE  Recommend one gold task from each arm. Set $r = 2$.

\LOOP
\STATE At the first $K$ time steps of the $r$-th epoch, recommend $K$ gold tasks, each from one arm. 
\STATE Choose arm $c_r = \argmax_{k} \bar{Y}_k(r)$.
\STATE Recommend a non-gold task drawn from arm $c_r$ at each of the following $\tau(r) - \tau(r-1)$ time steps. 
\STATE Set $r=r+1$.
\ENDLOOP
\end{algorithmic} 
\end{algorithm}

In this subsection, we propose another strategy, which we call the Uniform-Pulling Recommendation (UR) strategy. We again divide time into epochs, with the $r$-th epoch containing $K+\tau(r)-\tau(r-1)$ time steps. In the first $K$ time steps of the $r$-th epoch, where $r>1$, we first recommend $K$ gold tasks, one from each of the $K$ task categories. Next, we choose arm $c_r = \argmax_{1\le k\le K} \bar{Y}_k(T_k(r))$, where now $T_k(r)=r$ for all $k\in [1,K]$. In the subsequent $\tau(r)-\tau(r-1)$ time steps, we recommend a non-gold task from arm $c_r$ at each time step. The UR strategy is summarized in \cref{alg:UR}.

\begin{Theorem}\label{theorem:UR}
UR has regret of order $O(\sqrt{n})$, where $n$ is the number of time steps.
\end{Theorem}
\begin{IEEEproof}
The proof uses a similar argument and the same notations as that in the proof of \cref{theorem:GR}. For $r>1$, we have
\begin{align}
	P_{*,r} &\ge 1 - \sum_{k\ne *}\P(\bar{Y}_k(r)\ge \bar{Y}_*(r))\nonumber\\
	&= 1 - \sum_{k\ne *}\P(\bar{Y}_k(r)-\bar{Y}_*(r)+\Delta_k\ge \Delta_k)\nonumber\\
	&\ge 1- \sum_{k\ne *} e^{-2\Delta_k^2r} \nonumber\\
	&\ge 1- o(r^{-1}),\label{urcpr}
\end{align}
where the penultimate inequality follows from  Hoeffding's inequality.

We choose $M$ be the largest integer satisfying $ K+\sum_{r=2}^{M} \left( \tau(r) - \tau(r-1)+K \right) = \lceil\alpha M^{2}\rceil- \lceil\alpha^{2}\rceil + KM \le n$ so that $M = O(\sqrt{n})$, we have from \cref{rn,bbb},
\begin{align*} 
	&r(n) \nonumber\\ 
&\ge 	\sum_{r=2}^{M}q_*(2\alpha r-\alpha-1)^{+}\left(p_*P_{*,r} -  \E[\frac{\beta \sigma_{*}^{ 2}}{T_{*}(r)}]\right).\nonumber \\
&\ge \sum_{r=2}^{M}q_* p_*(2\alpha r-\alpha-1)^{+} \left( 1- o(r^{-1})\right) \nonumber \\
& \hspace{0.5cm} - \sum_{r=2}^{M}(2\alpha r-\alpha-1)^{+}\frac{\beta \sigma_{*}^{ 2}}{r} \nonumber \\
&\ge q_* p_* \alpha M^2 - O(M) - \sum_{r=2}^{M}(2\alpha r-\alpha-1)^{+}\frac{\beta \sigma_{*}^{ 2}}{r}\nonumber \\
&\ge nq_*p_* - O(\sqrt{n}).
\end{align*}
where the second inequality follows from \eqref{urcpr}.
Therefore, $R(n) \le O(\sqrt{n})$, and the proof of \cref{theorem:UR} is complete.
\end{IEEEproof}

\subsection{$\epsilon$-First Recommendation Strategy}\label{subsec:epsilon}

\begin{algorithm}[!ht]
\caption{$\epsilon$-First}
\label{alg:ER} 
\begin{algorithmic}[1] 
\REQUIRE $H = \lfloor\sqrt{n}\rfloor$.

\STATE Recommend $H$ gold tasks from each of the $K$ arms. 
\STATE Choose arm $c_H = \argmax_{k} \bar{Y}_k(H)$.
\STATE Recommend a non-gold task drawn from arm $c_H$ at each of the following $n-KH$ time steps. 
\end{algorithmic} 
\end{algorithm}

The $\epsilon$-first strategy was designed to deal with the budget-limited MAB problem \cite{tran2010epsilon}, where the budget refers to the total number of time steps. In the $\epsilon$-first strategy, the first $\epsilon$ fraction of the budget is used to explore the reward distributions of all arms, while the remaining $1-\epsilon$ fraction is used spent on the empirically best arm found in the initial exploration phase. To apply the $\epsilon$-first strategy to our task recommendation problem, we let $\epsilon = \Theta(1/\sqrt{n})$. Uniform pulling of arms is used in the exploration phase. The $\epsilon$-first strategy is summarized in \cref{alg:ER}, where $\epsilon = KH/n$.

\begin{Theorem}\label{theorem:ER}
If $\epsilon = \Theta(1/\sqrt{n})$, then the $\epsilon$-first strategy has regret of order $O(\sqrt{n})$, where $n$ is the number of time steps.
\end{Theorem}
\begin{IEEEproof}
Let $P_{k,H} = \P(c_H = k)$. Similar to \cref{urcpr} in the proof of \cref{theorem:UR}, it can be shown that $P_{*,H} \ge 1- o(H^{-1})$. From \cref{reward} and \cref{bbb}, we obtain
\begin{align}
&r(n) \nonumber\\ 
&\geq  \E[ q_{c_H}\left(n - KH\right)\left(p_{c_H} -  \frac{\beta\sigma_{c_H}^{ 2}}{Hq_{c_H}}\right)^{+}] \nonumber \\	
&\geq  \E[ q_{*}\left(n - KH\right)\left(p_{c_*} -  \frac{\beta\sigma_{*}^{ 2}}{Hq_{*}}\right)\indicator{c_H = *}] \nonumber \\	
&\geq  q_{*}\left(n - KH\right)\left(p_{*}P_{*,H} -  \frac{\beta\sigma_{*}^{ 2}}{Hq_{*}}\right)\nonumber \\
&\geq nq_*p_* - O(\sqrt{n}),	
\end{align}
since $H =O(\sqrt{n})$.
Therefore, $R(n) \le O(\sqrt{n})$, and the proof is complete.
\end{IEEEproof}

\subsection{Discussions}\label{subsec:gamma}

The $\epsilon$-first strategy assumes that the total number of tasks $n$ to be recommended to each worker is known beforehand, while the other two strategies do not need such an assumption. If both UR and the $\epsilon$-first strategy recommend the same total number of gold tasks, then the $\epsilon$-first strategy achieves a smaller regret than UR. This is because the $\epsilon$-first strategy recommends all the gold questions in the initial exploration phase, leading to a better estimation of the reward distributions before any non-gold tasks are recommended. However, in a practical crowdsourcing platform, knowing the total number of tasks for a worker may not be feasible as a worker is not guaranteed to be active or to remain in the system. An alternative is to use a hybrid UR and $\epsilon$-first strategy, where the initial exploration phase in each epoch of the UR strategy is set to be an $\epsilon$ fraction of the total number of time steps in that epoch.

On the other hand, UR can become more costly than GR when the number of arms $K$ is large. If UR and GR both recommend the same total number of gold tasks to the worker, GR tends to recommend more gold tasks from the best task category in the long run, leading to a smaller estimation variance of \cref{barX} for the best category and a smaller asymptotic regret than UR.

In both the GR and UR strategies, we choose $\tau(r)=\Theta(r^2)$, which determines how frequently gold tasks are recommended to the worker. A more general choice is $\tau(r)=\Theta(r^\gamma)$ for $\gamma \ge 1$. We now show that $\gamma=2$ is essentially the only choice that gives optimal order regret in GR and UR. Let $f(n)$ be the number of gold tasks recommended till time $n$. Then, since GR and UR recommends a fixed number of gold tasks in each epoch, we have $f(n)=\Theta(r)$ if $n$ is within epoch $r$. The number of time steps up to and including epoch $r$ is $\Theta(\tau(r))$, which implies that $f(n)=\Theta(n^{1/\gamma})$. Then from \cref{Rbdd1}, we have 
\begin{align*}
	R(n) &\ge \min \left\{ \frac{an}{b_1 n^{1/\gamma}}+ q_*p_* b_2 n^{1/\gamma} - a, nq_*p_*\right\}\\
	&=  \min \left\{ \frac{a}{b_1} n^{1-1/\gamma}+ q_*p_* b_2 n^{1/\gamma} - a, nq_*p_*\right\}  ,
\end{align*}
for some positive constants $a$, $b_1$, and $b_2$. Therefore, if $\gamma \ne 2$, $R(n)/\sqrt{n} \to\infty$ as $n\to\infty$.

\section{Simulations and Performance Evaluation}\label{sec:simulation}

In this section, we perform simulations and compare the performance of GR, UR, and the $\epsilon$-first strategy. We also compare with the UR strategy that uses $\tau(r)=\lceil \alpha r^{\gamma}\rceil$, for $\gamma > 0$, which we call the UR$(\gamma)$ strategy. The UR$(2)$ strategy is equivalent to the UR strategy. We perform simulations using different settings shown in \cref{tab:table1}.

\begin{table}[!htb]
\caption{$(p_k,q_k)$ for each task category. In setting no.~2, $(x,y)$ is varied over $[0,1]^2$.}
\label{tab:table1} 
\centering
        \small
        \setlength\tabcolsep{2pt}
\begin{tabular}{|c|c|c|c|c|}
 \hline
 \multirow{2}{*}{Setting no.} &\multicolumn{4}{c|}{Task category $k$} \\
 \cline{2-5}
  & 1 & 2 & 3 & 4 -- 10\\
\hline

	1 & $(0.7,0.7)$ & $(0.9,0.3)$ & $(0.3,0.9)$ & $(0.4,0.4)$\\
	2 & $(0.7,0.7)$ & $(x,y)$     &$(0.4,0.4)$ & $(0.4,0.4)$
\\ \hline
\end{tabular}

\begin{tabular}{|c|c|c|c|c|c|c|}
 \hline
 \hline
 \multirow{2}{*}{Setting no.} &\multicolumn{4}{c|}{Task category $k$} \\
 \cline{2-5}
  & 1 & 2 -- 10 & 11 -- 15 & 16 -- 25\\
\hline

	3 & $(0.8,0.8)$ &$(0.4,0.4)$  &   &\\
	4 & $(0.8,0.8)$ & $(0.4,0.4)$ & $(0.4,0.4)$&\\
	5 & $(0.8,0.8)$ & $(0.4,0.4)$ & $(0.4,0.4)$&$(0.4,0.4)$
\\ \hline
\end{tabular}
\end{table}

We track the regret as our performance measure. In each simulation, we perform $2000$ trials, each with $10^3$ time steps. To compare the performance of different approaches, we compute the empirical average regret, over all the trials. For convenience, this average is referred to as the \emph{average regret}.


In \cref{theorem:GR}, we show that for GR, a sufficient condition, under which the order optimal regret is achieved, is $c > \max\{5d^2, 2\}$. However, in simulations we find that a smaller $c$ can lead to better performance for all settings listed in \cref{tab:table1}. Therefore, in the simulations, we set $c = 0.05$ in GR. In practice, the parameter $d$ in GR can be set to be a fixed small value like $0.1$. We set the parameter $d$ in GR to be $0.1$. In all simulations except those in \cref{subsec:vary_alpha}, we set $\alpha=0.1$ for GR, UR and UR$(\gamma)$ for all $\gamma$. We use $\beta=10$ in all simulations.

\subsection{Comparison of Different Strategies}
\label{subsec:differ_stra}

\begin{figure}[!htb]
  \centering
\includegraphics[width=1\linewidth]{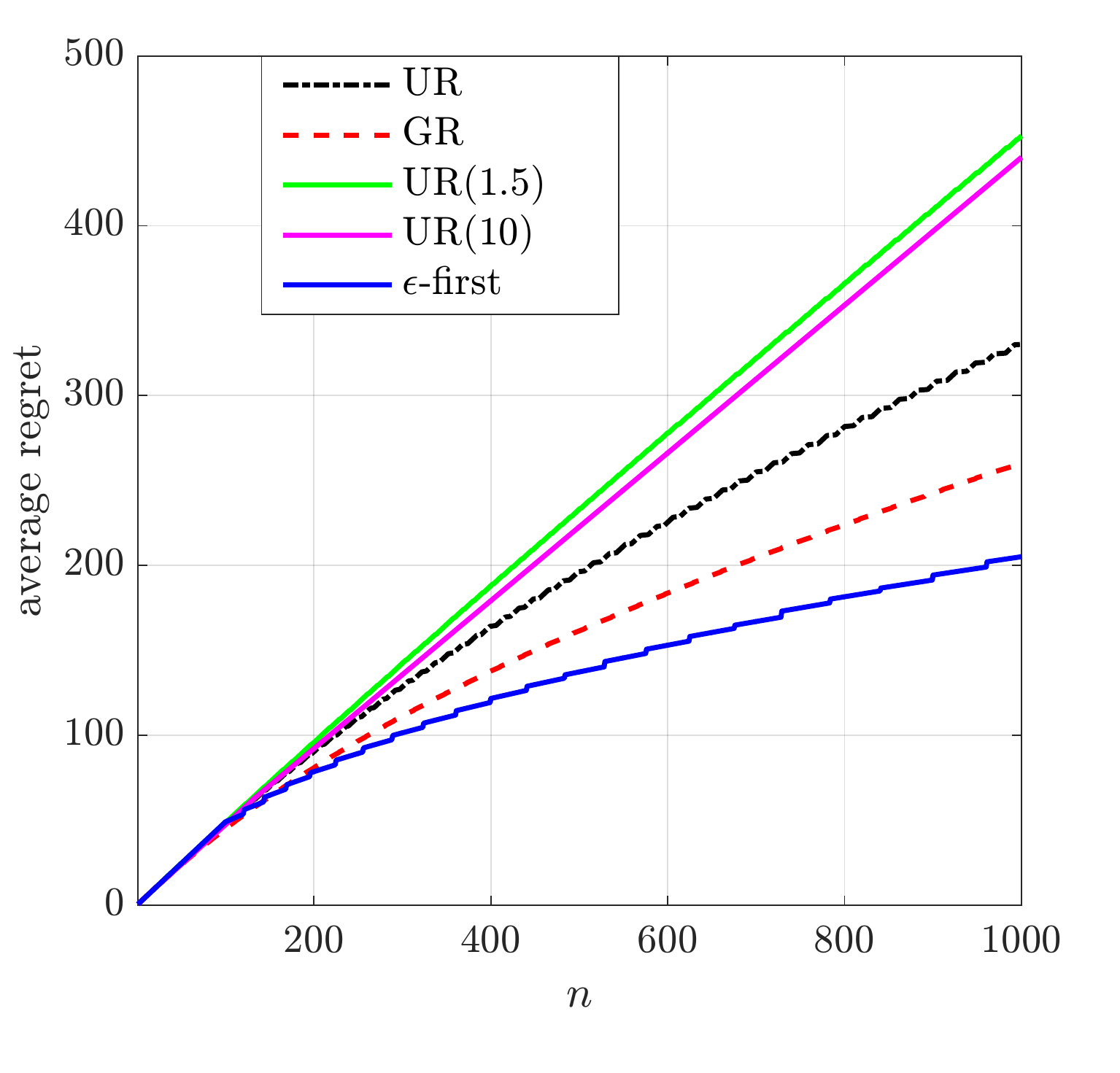}
 \centering
\caption{Performance comparison between different strategies using setting no.~1.}
\label{fig:D1}
\end{figure}

\begin{figure}[!htb]
  \centering
\includegraphics[width=1\linewidth]{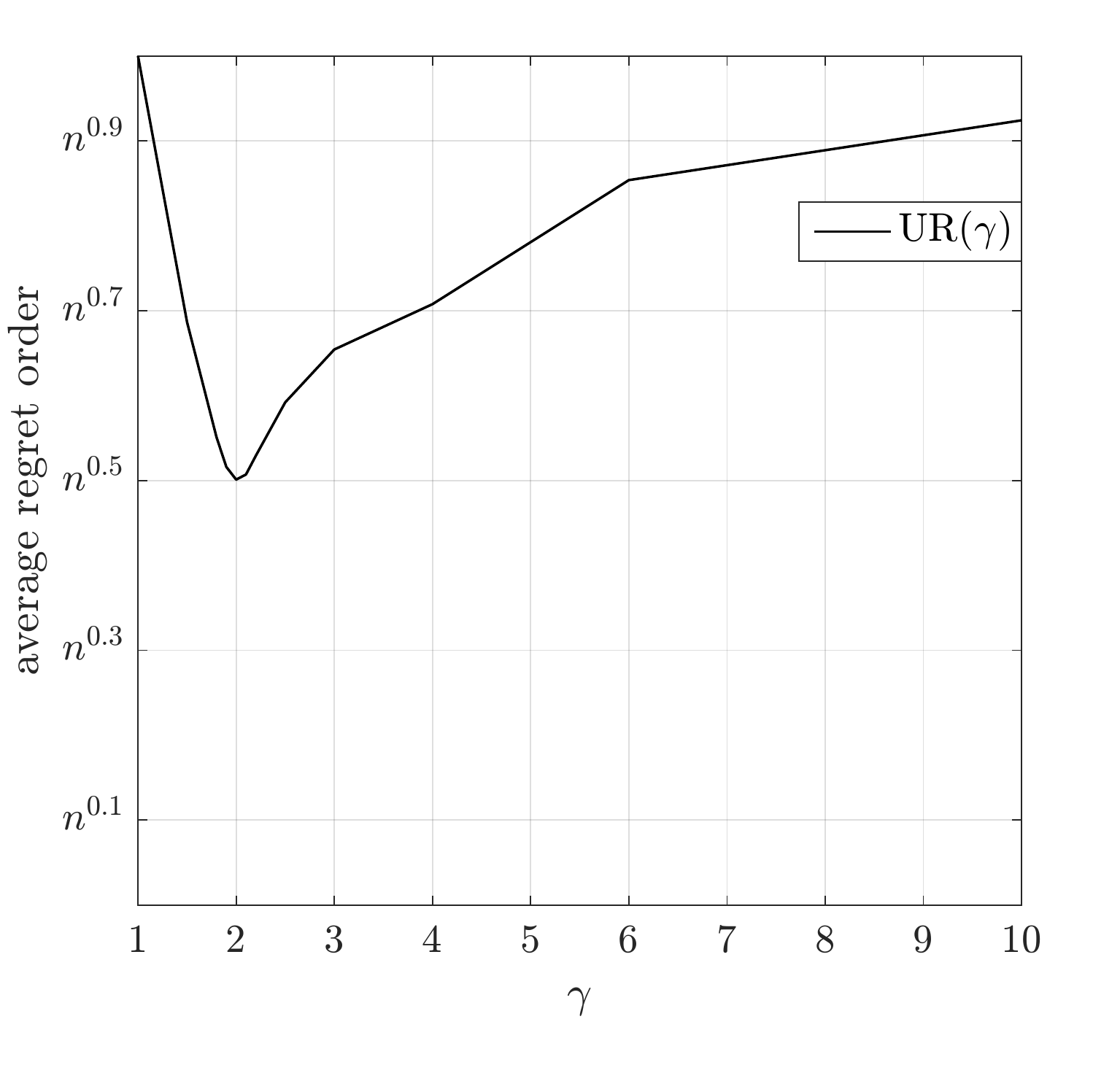}
 \centering
\caption{Average regret order of UR$(\gamma)$ using  setting no.~1.}
\label{fig:gamma} 
\end{figure}

We first compare the performance of different strategies.  In \cref{fig:D1,fig:gamma}, we use setting no.~1 in \cref{tab:table1} for our simulations. We observe the following.
\begin{itemize}

\item GR, UR, and the $\epsilon$-first strategy outperform UR$(1.5)$ and UR$(10)$ when $n$ is sufficiently large. In UR$(1.5)$, gold tasks are recommended more frequently than GR and UR. Therefore, less non-gold tasks are recommended to the worker. On the other hand, UR$(10)$ corresponds to the case where gold tasks are recommended infrequently compared to GR and UR. In \cref{fig:gamma}, we see that $\gamma=2$ achieves the optimal regret order, in line with our observations in \cref{subsec:gamma}.

\item The $\epsilon$-first strategy performs best among all the strategies, but it requires to know the total number of time steps $n$ beforehand.
\end{itemize}

\subsection{Varying Total Number of Task Categories \texorpdfstring{$K$}{K}}
\begin{figure}[!htb]
\centering
  \includegraphics[width=1\linewidth]{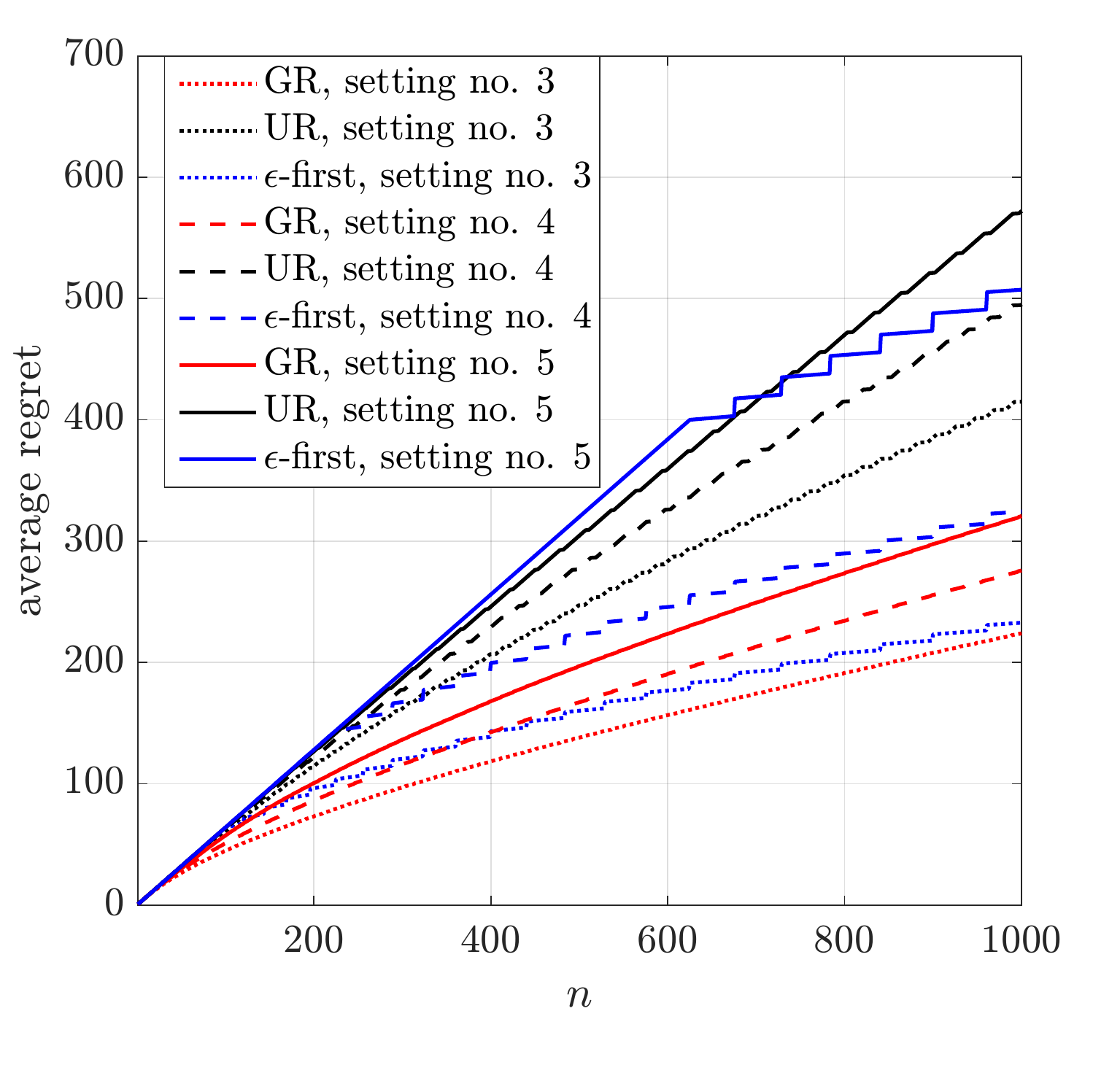}
  \centering
\caption{Performance comparison between UR and GR with $K = 10$, $15$ and $25$ using setting no.\ 3 -- 5 in \cref{tab:table1}, respectively.}
  \label{fig:Dcombine}
\end{figure}

We next show the performance of UR, GR and the $\epsilon$-first strategy for different number of arms $K$, using the settings no.~3 -- 5 listed in \cref{tab:table1}. From \cref{fig:Dcombine}, we observe the following.
\begin{itemize}
\item  GR outperforms UR. One reason is the exploration of all arms at every epoch in UR reduces the cumulative reward with fewer non-gold tasks assigned. Another reason is even if GR and UR have recommended the same total number of gold tasks to the worker, the number of gold tasks recommended from the best category using UR is less than that using GR.  Therefore, the variance of the empirical mean \cref{barX} for the best category using UR is larger than that using GR, leading to a larger average regret.
\item The average regret with more task categories is larger than that with fewer task categories, which is expected.
\end{itemize}

\subsection{Varying \texorpdfstring{$\alpha$}{alpha}}\label{subsec:vary_alpha}

\begin{figure}[htb]
  \centering
\includegraphics[width=1\linewidth]{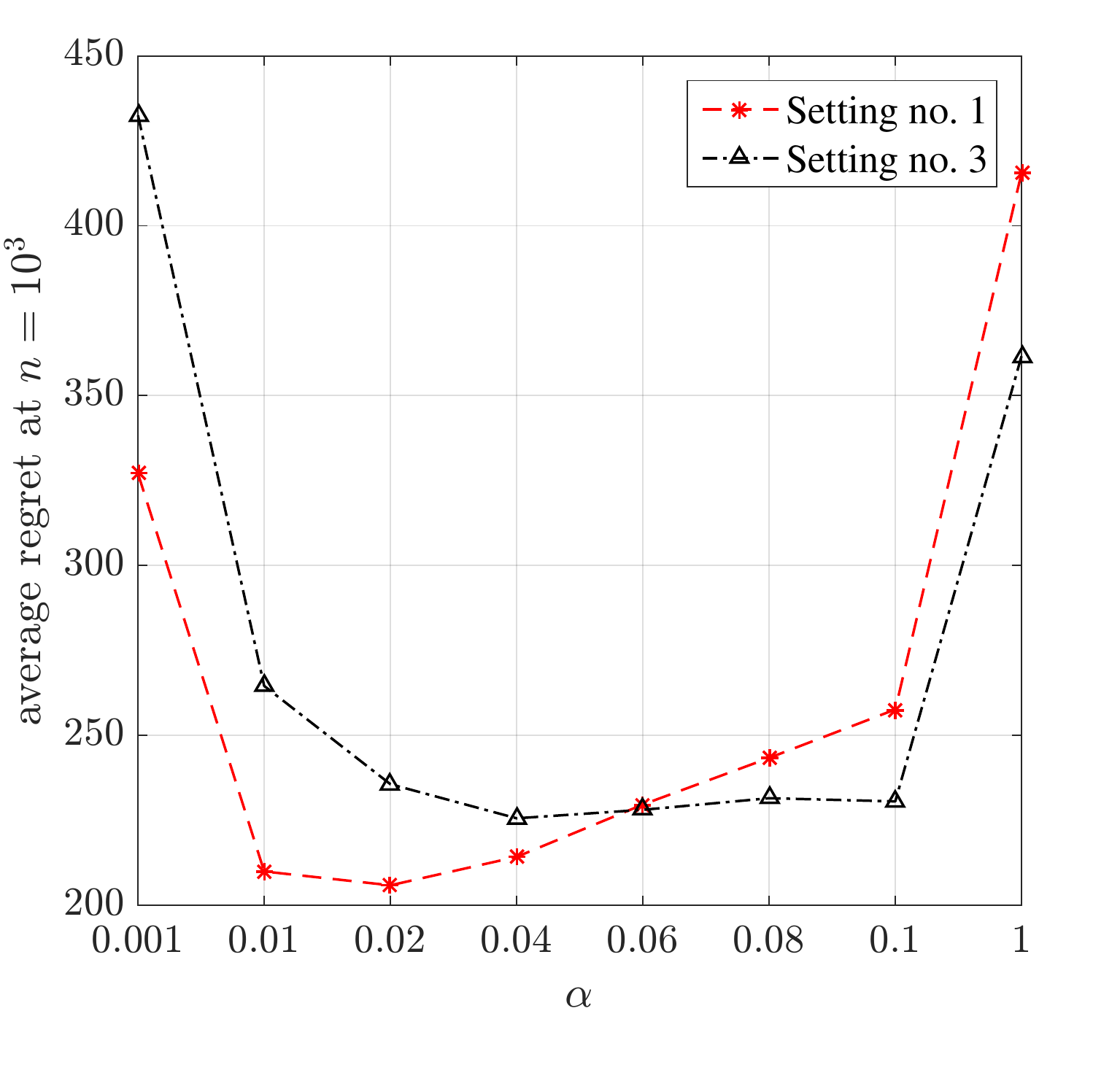}
 \centering
\caption{Performance of GR with different values of parameter $\alpha$ and different settings.}
\label{fig:alpha_GR}
\end{figure}

\begin{figure}[!htb]
\centering
  \includegraphics[width=1\linewidth]{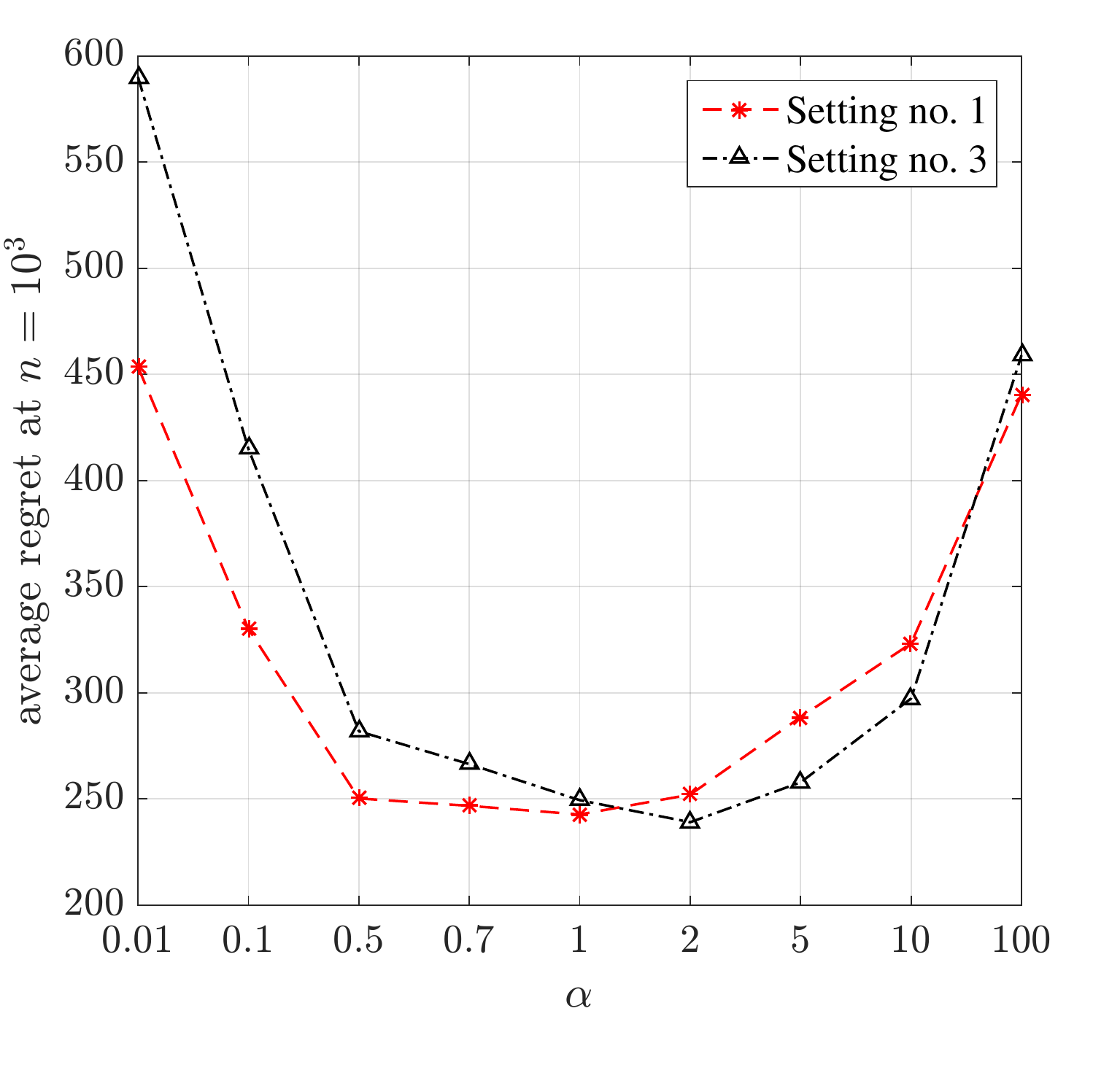}
  \centering
\caption{Performance of UR with different values of parameter $\alpha$ and different settings.}
  \label{fig:alpha_UR}
\end{figure}

We next vary the parameter $\alpha$ in GR and UR separately using setting no.~1 and 3. The results are shown in \cref{fig:alpha_UR,fig:alpha_GR}, respectively. We observe the following:
\begin{itemize}
\item We note that a moderate $\alpha$ is optimal for both UR and GR. When $\alpha$ is too large, the strategies recommend gold tasks too infrequently, leading to a larger estimation variance. On the other hand, a small $\alpha$ results in too many gold tasks.
\item The value $\min_{k\ne *} \Delta_k$ in setting no.~3 is larger than that in setting no.~1. The optimal $\alpha$ in both GR and UR tends to decrease with decreasing $\min_{k\ne *} \Delta_k$. This is because when $\min_{k\ne *} \Delta_k$ decreases, increasing the number of gold questions allows for greater exploration of the task categories.
\end{itemize}


\subsection{Varying \texorpdfstring{$\min_{k\ne *}\Delta_k$}{Deltak}} 
\label{subsec:vary_d}
\begin{figure}[!htb]
\centering
  \includegraphics[width=1\linewidth]{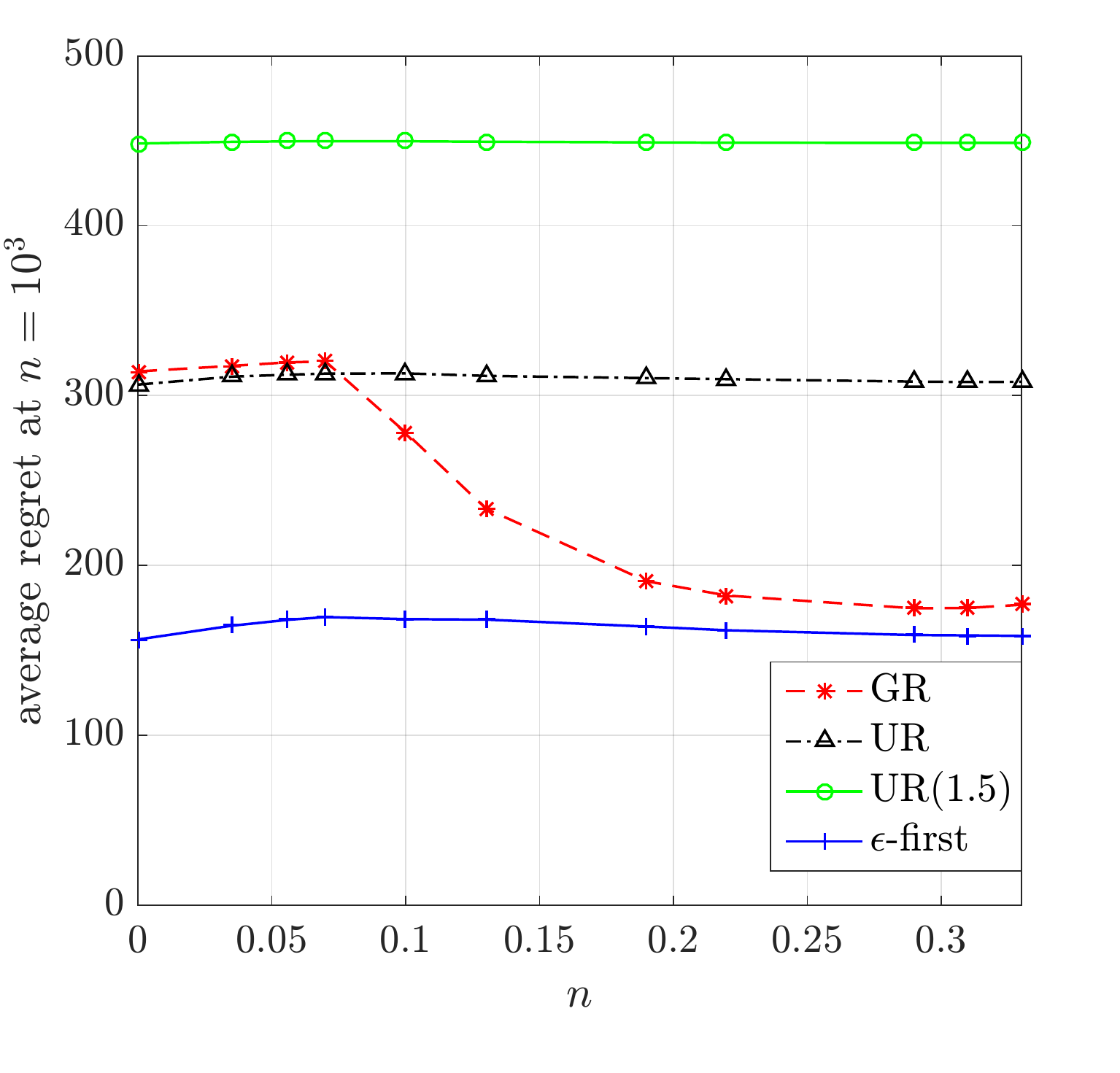}
  \centering
\caption{Performance of different strategies with varying $\min_{k\ne *} \Delta_k$ using setting no.~2 for different values of $(x,y)$.}
  \label{fig:delta}
\end{figure}

When $\min_{k\ne *} \Delta_k$ is small, distinguishing the best task category $*$ from the other categories is difficult. We vary the values $(x,y)$ in setting no.~2 in \cref{tab:table1} to obtain different values of $\min_{k\ne *} \Delta_k = q_*p_* - xy = 0.49-xy$. The results are shown in \cref{fig:delta}, where we plot the average regret at the time step $n = 10^3$. We observe that a relatively smaller $\min_{k\ne *} \Delta_k$ results in a larger average regret for GR. This is because a smaller $\min_{k\ne *} \Delta_k$ makes it more difficult to distinguish between the best arm and the others. We also observe that the average regrets for UR and the $\epsilon$-first strategy remain almost constant with varying $\min_{k\ne *} \Delta_k$ because these strategies use a sufficiently large number of gold tasks to estimate the parameters of each arm.

\begin{figure}[!htb]
\centering
  \includegraphics[width=1\linewidth]{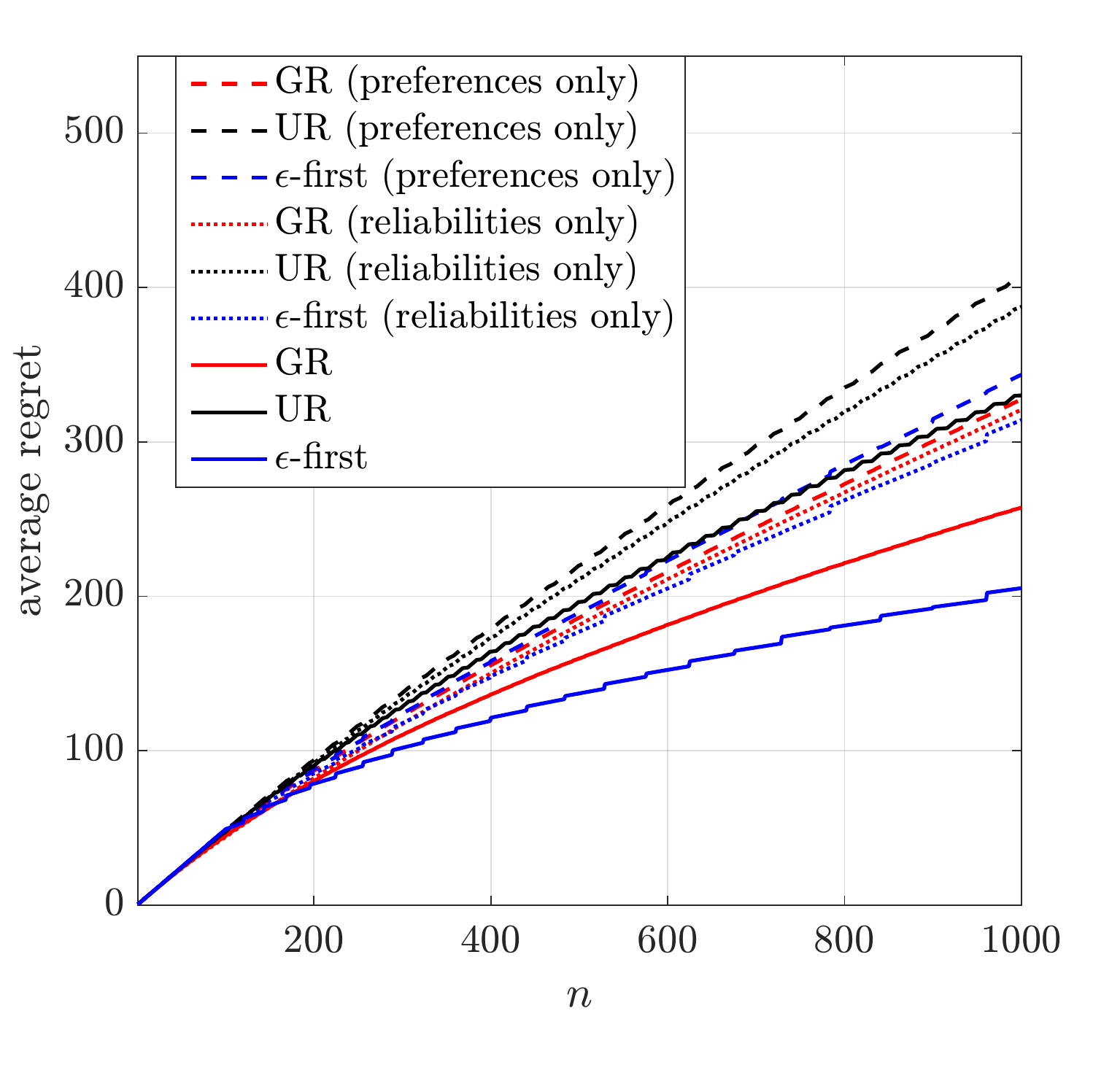}
  \centering
\caption{Performance of UR, GR and the $\epsilon$-first strategy considering either preferences or reliabilities under setting no.~1.}
  \label{fig:partial_infor}
\end{figure}

\subsection{Learning Only Partial Information} 
\label{subsec:partial}

We next consider a crowdsourcing platform that learns only either a worker's preferences or reliabilities but not both, when performing task recommendations. The simulation results are shown in \cref{fig:partial_infor} using setting no.~1 in \cref{tab:table1}. We see that learning either only the preferences or reliabilities results in a lower cumulative reward, which is expected. The loss in cumulative reward for GR, UR or the $\epsilon$-first strategy learning either only the preferences or reliabilities widens compared to GR, UR or the $\epsilon$-first strategy respectively when the number of time steps $n$ becomes large.

\section{Conclusion} \label{sec:conclusion}

We have formulated the task recommendation problem in crowdsourcing as a MAB problem in which the reward is unobservable. To learn the worker's preferences and reliabilities for different task categories, we have included gold tasks whose solutions are known \emph{a priori}, but which however do not produce any rewards. We showed that the optimal regret order is $O(\sqrt{n})$, where $n$ is the number of time steps. We proposed three strategies and showed that all achieve the optimal regret order. Simulations verify the efficiency of our proposed strategies.  

In this paper, we have assumed that the worker's preferences and reliabilities for different task categories are independent. However, in practice, preferences and reliabilities for different categories may be dependent, and such correlation information can be used to enhance the exploration phase. A future direction is to consider the case where workers' reliabilities and preferences for different task categories may be correlated, and to study how to recommend tasks from different categories to multiple workers.

\bibliographystyle{IEEEtran}
\bibliography{IEEEabrv,refrecomm}

\begin{IEEEbiography}{Qiyu Kang}(S'17) received the B.S. degree in Electronic Information Science and Technology from University of Science and Technology of China in 2015. He is currently a Ph.D. candidate at the School of Electrical and Electronic Engineering, Nanyang Technological University. His research interests include distributed signal processing, collaborative computing, and social network.
\end{IEEEbiography}

\begin{IEEEbiography}{Wee Peng Tay}(S'06 M'08 SM'14) received the B.S. degree in Electrical Engineering and Mathematics, and the M.S. degree in Electrical Engineering from Stanford University, Stanford, CA, USA, in 2002. He received the Ph.D. degree in Electrical Engineering and Computer Science from the Massachusetts Institute of Technology, Cambridge, MA, USA, in 2008. He is currently an Associate Professor in the School of Electrical and Electronic Engineering at Nanyang Technological University, Singapore. His research interests include distributed inference and signal processing, sensor networks, social networks, information theory, and applied probability.

Dr. Tay received the Singapore Technologies Scholarship in 1998, the Stanford University President's Award in 1999, the Frederick Emmons Terman Engineering Scholastic Award in 2002, and the Tan Chin Tuan Exchange Fellowship in 2015. He is a coauthor of the best student paper award at the Asilomar conference on Signals, Systems, and Computers in 2012, and the IEEE Signal Processing Society Young Author Best Paper Award in 2016. He is currently an Associate Editor for the IEEE Transactions on Signal Processing, an Editor for the IEEE Transactions on Wireless Communications, serves on the MLSP TC of the IEEE Signal Processing Society, and is the chair of DSNIG in IEEE MMTC. He has also served as a technical program committee member for various international conferences.
\end{IEEEbiography}
\end{document}